\documentclass{aa}
\usepackage{graphics}
\begin{document}

\thesaurus{}

\def\EE#1{\times 10^{#1}}
\def\gcm{\rm ~g~cm^{-3}}
\def\cm3{\rm ~cm^{-3}}
\def\kms{\rm ~km~s^{-1}}
\def\cms{\rm ~cm~s^{-1}}
\def\ergs{\rm ~erg~s^{-1}}
\def\wl{~\lambda}
\def\wll{~\lambda\lambda}
\def\FeI{{\rm Fe\,I}}
\def\FeII{{\rm Fe\,II}}
\def\FeIII{{\rm Fe\,III}}
\def\Msun{~{\rm M}_\odot}

\def\lsim{\!\!\!\phantom{\le}\smash{\buildrel{}\over
  {\lower2.5dd\hbox{$\buildrel{\lower2dd\hbox{$\displaystyle<$}}\over
                               \sim$}}}\,\,}
\def\gsim{\!\!\!\phantom{\ge}\smash{\buildrel{}\over
  {\lower2.5dd\hbox{$\buildrel{\lower2dd\hbox{$\displaystyle>$}}\over
                               \sim$}}}\,\,}

\title{Deep optical observations at the position of PSR1706-44 with the VLT-UT1\thanks{Based on
observations collected at the European Southern Observatory, Paranal,
Chile (VLT-UT1 Science Verification Program)}}

\author{
Peter Lundqvist\inst{1}
\and Jesper Sollerman\inst{1} 
\and Alak Ray\inst{2} 
\and Bruno Leibundgut\inst{3} 
\and Firoza Sutaria\inst{4}
}
\institute{
Stockholm Observatory, SE-133 36 Saltsj\"obaden, Sweden
\and Tata Institute of Fundamental Research, Bombay 400 005, India
\and European Southern Observatory, 85748 Garching bei M\"unchen, Germany
\and Inter University Centre for Astronomy \& Astrophysics, Pune 411 007, India
}

\date{Received:  Accepted: }

\mail{peter@astro.su.se}

\titlerunning{Optical observations of PSR1706-44 with VLT}
\maketitle

\begin{abstract}

We present optical data gathered by the {\it VLT} Test Camera in 
the {\it V}-band at the radio (interferometric) position of PSR1706-44. We
find no optical counterpart to the pulsar in the {\it VLT} image. At a 
distance of 2\farcs7 from a nearby bright star, the 3$\sigma$ upper limit 
to the pulsar magnitude above the background is $V = 25.5$. Within an error
circle of 1\farcs0 the upper limit is degraded in the direction
towards the star. At a distance $\lsim 2\arcsec$ from the star we can with
confidence only claim an upper limit of $V = 24.5$. This is still several
magnitudes fainter than previous estimates. The implications of the optical
upper limit taken together with the high energy pulsed gamma-ray radiation 
for theoretical models of pulsar emission are discussed. 

\end{abstract}



\section{Introduction}

Optical detection of pulsars
constitute a critical part of an expanding set of multiwavelength observations
of isolated neutron stars that together aid in the development
and constraining of theoretical models of pulsar electromagnetic radiation. 
To detect optical pulsations, it is necessary to unambiguously
identify the optical counterpart of a pulsar that has been observed in
other bands, say radio or gamma-rays.
Because of its high spin-down energy loss and relative proximity to Earth, 
the radio pulsar PSR1706-44 has been a prime candidate for observation in 
many bands of the electromagnetic spectrum. (See Table 1 for a description of
PSR1706-44, and a comparison with the Vela pulsar, which is of similar age.)
The recent observation by the {\it Very Large Telescope (VLT)}-UT1 in its 
Science Verification phase, of the field containing this pulsar has allowed 
the determination of a magnitude limit in the $V$-band.
 
\begin{table}
\caption{Properties of PSR1706-44 and PSR0833-45 (Vela)$^{a}$}
\label{properties}

\[
\begin{tabular}{lcc}
\hline

\noalign{\smallskip}

 & PSR1706-44 & Vela pulsar\\

\noalign{\smallskip}
\hline
\noalign{\smallskip}

Distance (kpc), $d$                              & 1.8$^{b}$  & 0.5    \\
Pulse period, $P$ (s)                            & 0.1024     & 0.0893 \\
$\dot P$ ($10^{-15}$~s s$^{-1}$)                 & 93.04      & 124.7  \\
log (Timing age)                                 & 4.24       & 4.05   \\

\noalign{\smallskip}
\hline
\end{tabular}
\]

$^{a}$ Data from Taylor, Manchester \& Lyne (1993).\\
$^{b}$ Alternate $d=2.4$~kpc (Koribalski et al. 1995; Thompson et al. 1996).

\end{table}

A number of rotation powered pulsars are found to emit high energy
radiation (from optical to gamma-ray bands) which is a combination of
differing amounts of three spectral components: 1)  power-law emission,
resulting from non-thermal radiation of particles
accelerated in the pulsar magnetosphere, 2) soft blackbody emission
from surface cooling of the neutron star, 3) a hard thermal
component from heated polar caps. In addition, there
is often a background of unpulsed emission from a surrounding 
synchrotron nebula. 

PSR1706-44 belongs to the set of seven $\gamma$-ray pulsars detected by 
EGRET (Thompson et al. 1996). It 
has been detected as an unpulsed point source by ROSAT (Becker et al. 1995).
While it has not yet been seen as a pulsed X-ray source, strong
upper limits to its pulsed X-ray flux from the Rossi X-ray Timing Explorer
and other satellites
has been used to constrain the level of the thermal component from the
heated polar caps. In the optical, PSR1706-44 has not been detected. Deep
optical observations like that in this work are needed to meaningfully
test the outer gap model's prediction of optical emission (Ray et al. 1999).

At present, it is not clear how and where in the pulsar magnetosphere
the pulsed non-thermal high energy emission originates. Similarly, the
relationship between optical pulsed emission and those in the X-ray
or gamma-ray bands are unclear. Qualitatively, high energy radiation
is believed to occur from incoherent curvature radiation in the outer
magnetosphere or by synchrotron emission by energetic electrons near
the light cylinder. So far optical pulsations have been detected from
the Crab and Vela pulsars, PSR0540-69, PSR0656+14 and (possibly) Geminga, 
while ultraviolet pulsations were seen only from the Crab
pulsar using the Hubble Space Telescope (Gull et al. 1998).
The existing models of optical pulsed radiation (Pacini \& Salvati 1987)
underpredict the observed fluxes of middle aged pulsars like Geminga
and PSR0656+14 by several orders of magnitude. A phenomenological
analysis of the optical efficiencies (fraction $\eta_{\rm opt}$ of spin down
power radiated in the optical bands) on pulsar parameters show that
$\eta_{\rm opt} \propto \dot P^2$ for the five observed so far
(Goldoni et al. 1995). However, the overall consistency of the
models with observed data like phase relationship and the correlation
between optical and higher energy bands is not very compelling.

Here we report on the {\it VLT}-UT1 Science Verification phase observations 
in the optical of the field including the position of the radio emission from 
PSR1706-44. The results are discussed in Sect. 2, and their implications in 
Sect. 3.

\section{Observations and Results}

The field of PSR1706-44 was observed with the Test Camera on {\it VLT}-UT1 on
August 19, 1998, during Science Verification.
Six images of 600 seconds each were obtained in the $V-$band. All
observations were made with 2x2 binning, a pixel thus corresponds to
$\sim 0\farcs09$ on the sky. The raw images were bias subtracted by
determining the bias level in the overscan region of the CCD.
The two-dimensional bias structure was removed with a master bias
frame. Flatfielding was done using a $V-$flat obtained from the science
observations on the previous night\footnotemark.
The six images were aligned and combined into a final image (see Fig.~1).
The quality of this image is very good, with a FWHM of $\sim 0\farcs5$.
\footnotetext{See http://www.eso.org/paranal/sv/ for details.}

Previous attempts to constrain the emission from the pulsar have been severely
hampered by a bright nearby star. This star was named Star 1 by Chakrabarty \& 
Kaspi (1998; henceforth CK98), and its magnitude was measured to $V = 17.3$.
Due to poorer spatial resolution they could only obtain an upper limit for
the pulsar, $R = 18$. The good seeing of the {\it VLT} image enables us to
significantly improve upon this.

\begin{figure}
\resizebox{\hsize}{!}{\includegraphics{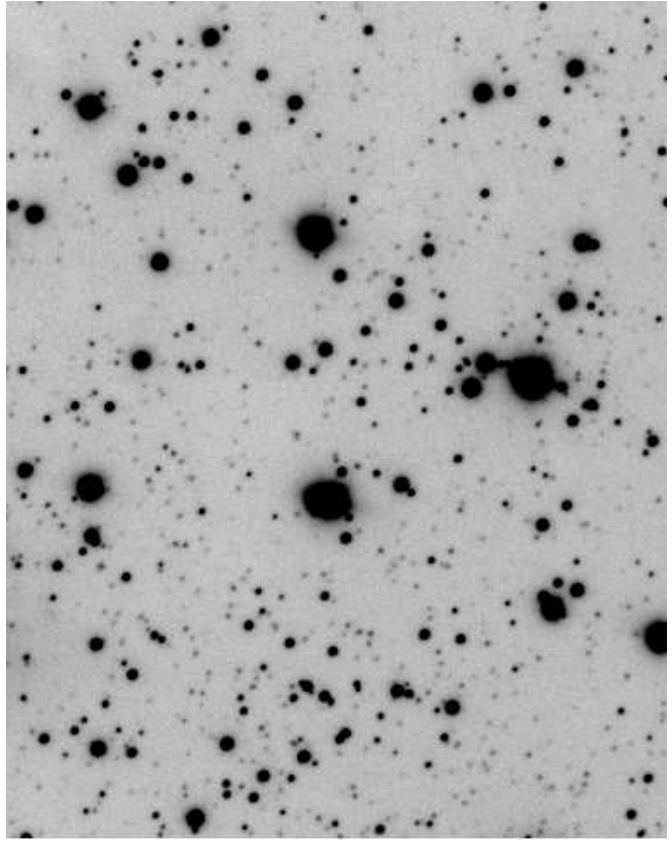}}
\caption{
The field of PSR1706-44 as observed with the Test Camera on {\it VLT}-UT1.
This is the combination of six {\it V}-images with a total exposure time of 
3600s. The field of view shown is 70\arcsec x 88\arcsec. The image is rotated
45\degr\ clockwise. The position of the pulsar is to the far left of the image
(see Fig.~2).
}
\label{fig1.ps}
\end{figure}

\begin{figure}
\resizebox{\hsize}{!}{\includegraphics{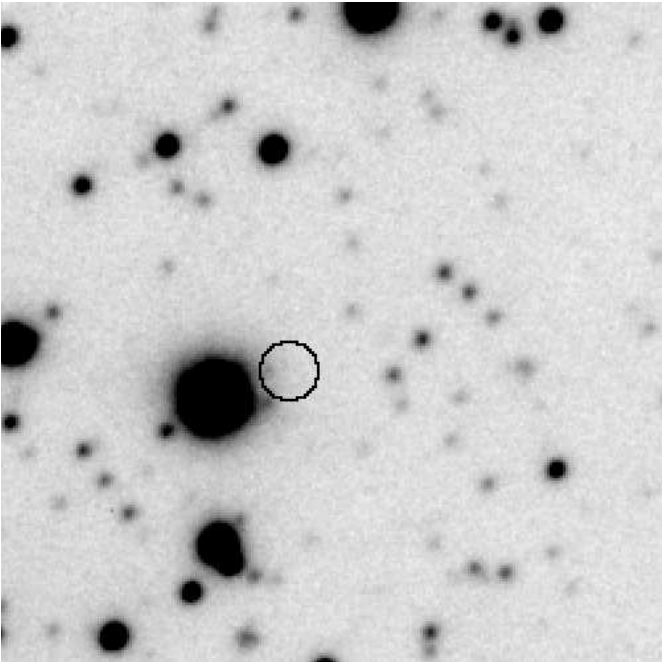}}
\caption{
This is a blow-up of Fig.~1 around Star 1. The expected position of the
pulsar is inside the error circle shown. The circle has a radius of 1\farcs0 
and is centered 2\farcs7 away from Star 1.
}
\label{fig2.ps}
\end{figure}

In Fig.~2 we show a blow-up of the region around Star 1.
Using the radio position of the pulsar (Frail \& Goss 1998; Wang et al. 1998),
CK98 estimate that the pulsar should lie $2\farcs7$ away from the star.
The uncertainty in this position is a combination of errors in the radio
position, errors in the astrometric solution to the optical image and a
mismatch in aligning radio and optical frames. These are all of the order
$0\farcs5-0\farcs7$. We have adopted a combined error of $1\farcs0$.
This error circle around the position $2\farcs7$ away from Star 1 is
shown in Fig.~2.

We carefully searched for the pulsar around that position in the {\it VLT}
image, but no object was found. To estimate an upper limit to the pulsar's
emission we constructed a PSF from ten bright stars in the field using the
IRAF/DAOPHOT PSF task. We thereafter added a number of artificial stars with
different magnitudes to the image. The magnitudes are all measured
relative to Star 1 ($V = 17.3$, CK98). The colour terms for the configuration
have been measured to be small and are neglected in our study.

To ensure similar backgrounds, the artificial stars were all positioned at a
distance of $2\farcs7$ from Star 1. We thus find that stars
of magnitude $V=25.0$ should have been easily seen. Also $V=25.5$
is clearly visible but an artificial star with $V=26.0$ is rather faint.
If the pulsar is positioned at the outer end of our error circle, even
a $V=26.0$ star is easily seen. Measuring the background at this distance
from Star 1 shows that an artificial star with $V=25.5$ has in fact a peak
pixel value that is more than 3$\sigma$ above the background. This is thus a
firm upper limit for the image.

However, if the pulsar would be positioned much closer to the star than
$2\farcs7$, our method of measuring the limiting magnitude becomes more
uncertain. In fact, we do see a region of brighter emission in the
innermost part of our error circle. This might just be fluctuations in the
PSF of the bright star. To estimate how bright a star one could hide in the PSF
of Star 1 we instead subtracted artificial stars from this position until a
hole appeared in the background. We find that it is possible to hide a rather
bright star ($V=25.0$) at a distance of $\lsim 2\arcsec$ from the star. As a
firm upper limit for a pulsar this close to Star 1 we therefore
claim $V=24.5$.

We conclude that the pulsar is most likely fainter than V=26.0 magnitudes. 
Deeper exposures are needed to address this question. However, if 
the pulsar is indeed substantially closer to the bright Star 1,
the PSF of that star limits our study. We claim an upper limit of
V=24.5 inside our $1\farcs0$ error circle. This is still much fainter
than the previous upper limit of CK98. HST resolution would be required to
improve upon this estimate.

\section{Discussion and Conclusion}

The magnitude limits obtained in Sect. 2 have theoretical implications which
we briefly mention here. The optical radiation is expected to be produced
by tertiary e$^{\pm}$-pairs produced in outer gap discharges. These
particle fluxes and energy spectra are in turn dependent upon those
of the primary and secondary electrons and the particular radiation
mechanism involved, and the optical fluxes may be correlated with the 
gamma-ray photon fluxes in such models.

\begin{figure}
\resizebox{\hsize}{!}{\includegraphics{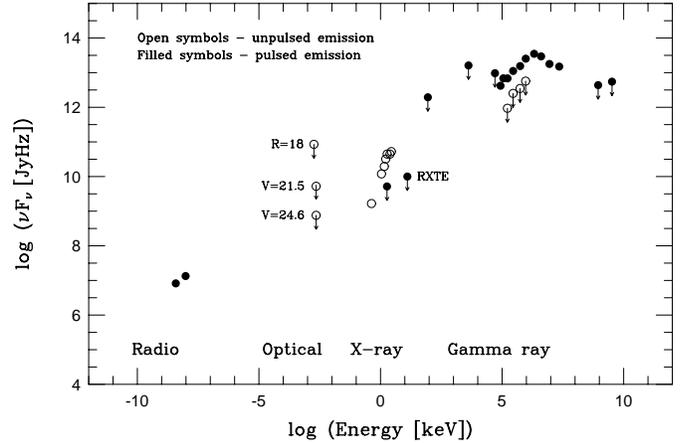}}
\caption{
Multiwavelength spectrum, $\nu F_{\nu}$ versus $\nu$, for PSR1706-44. Data
are taken from the compilations of Thompson et al. (1996, 1999), except for
the points marked `V' which are from this work. The point marked `RXTE' is
from Ray et al. (1999), and that marked `R' is from CK98. Our two V points 
correspond to two estimates which should bracket the dereddened upper limit 
in the visual: $V=24.6$ assumes minimum extinction ($A_{\rm V} = 0.9$) and 
the pulsar lying outside the PSF of Star 1, $V=21.5$ assumses maximum 
extinction ($A_{\rm V} = 3$) and the pulsar lying close to Star 1 in 
projection. The point by CK98 is likely to underestimate the extinction.
}
\label{fig3.ps}
\end{figure}

Usov (1994, see his Eqn.~[24]) has estimated the scaling of the optical vs
gamma-ray luminosities expected in the outer gap models by Cheng, Ho \& 
Ruderman (1986a, 1986b) for Vela-like pulsars. Usov's analysis predicts that 
the frequency integrated optical flux (between $4\EE{14}$ and $7.5\EE{14}$ Hz), 
$F_{\rm opt}$, should scale with the integrated gamma-ray flux 
as $F_{\rm opt} \gsim 4 \times 10^{-4} F_{\gamma}$. From Fig. 3 we estimate
the pulsed $F_{\gamma}$ to be $3 \times 10^{13}$ JyHz for PSR1706-44.
For a flat optical spectrum, this gives $F_{\nu} \gsim 3.4\EE{-28}$ erg 
s$^{-1}$ cm$^{-2}$ Hz$^{-1}$ between $4\EE{14}$ and $7.5\EE{14}$ Hz. Assuming
$F_{\nu}$ to be the same in both the $V$ and $R$ bands, the magnitudes
predicted by the outer gap models by Cheng, Ho \& Ruderman (according to Usov
1994) are $R \lsim 20.0$ and $V \lsim 19.8$. We will now compare these limits
with the observed limits by CK98 and those found in this work.

CK98 found an upper limit to the $R$ magnitude for the pulsar of $R = 18$.
It is obvious that the data of CK98 do not constrain the outer gap model, 
in particular since extinction appears not to have been included by CK98.

We know that extinction in the direction to the pulsar must occur.
The photometry of Star 1 by CK98 indicates that $E(B-V) \sim 0.29$. The 
extinction to the more distant PSR1706-44 is likely to be higher. A rough
estimate of the visual extinction is one magnitude per kpc (Spitzer 1978),
which would indicate $A_{\rm V} \sim 2$ to the pulsar. This is consistent
with the column density, $N_{\rm H} \lsim 5\EE{21}$ cm$^{-2}$, indicated by the
X-ray data of Finley et al. (1998). This limiting column density translates 
into $A_{\rm V} \lsim 3$. A likely range for $A_{\rm V}$ is therefore $0.9 - 3$
magnitudes. The dereddened $R$ magnitude could therefore be much brighter 
than $R = 18$, and of little value in constraining the outer gap model for 
PSR1706-44.

The situation is different for our $V$ estimates from the {\it VLT} 
observations. Even if we adopt maximum extinction ($A_{\rm V} \sim 3$), and if
the pulsar would lie close to Star 1 in projection, the dereddened observed
upper limit is only $V \simeq 21.5$, which is $\sim 1.7$ magnitudes fainter
than the limit obtained from Usov's analysis. In Fig.~3 we have included our
dereddened upper limit of $V$ (for two combinations of $A_{\rm V}$ and 
projected distances from Star 1) in a multiwavelength spectrum of the pulsar. 
(Similar spectra are presented by Thompson et al. (1999) for the Crab, Vela and
Geminga pulsars, as well as PSR1509-58, PSR1951+32 and PSR1055-52.) Our faint
limit to the $V$ flux therefore calls for a revision of the standard outer gap
model for PSR1706-44.

What revision could be inferred from the {\it VLT} observations? 
In the prediction of the ratio of $F_{\rm opt} / F_{\gamma}$ for 
Vela-like pulsars like PSR1706-44, an important assumption is that the gap
averaged magnetic field $\bar B$ is approximately equal to the magnetic
field at the outer boundary of the outer gap. This average $\bar B$ is:
$B_{\rm L} \leq \bar B \leq (c/\Omega r_{\rm i})^3 B_{\rm L}$, where 
the subscript `L' refers to the field at the light cylinder, $r_{\rm i}$ 
being the inner radius of the outer gap, and $\Omega = 2\pi / P $ is the spin 
frequency of
the pulsar. For small inclination angle $\chi$ between the magnetic moment 
and spin vectors one has $r_{\rm i} \Omega /c = 2/3$, 
and $\bar B \leq (3/2)^3 B_{\rm L}$.
This gives the synchrotron cutoff frequency $\nu_{\rm s}$ for tertiary photons
near $\sim 10^{14}$ Hz, so that optical emission from a pulsar active in the
gamma-ray region should be observable (as estimated above). On the other hand, 
if $\chi \geq \pi/4$, the synchrotron cutoff frequency is $\sim 10^{13}$ Hz 
and in this case the flux of
optical band radiation may be very small. Our faint limit from the {\it VLT}
for PSR1706-44 in combination with the outer gap model could therefore point 
to a case of an unaligned rotating neutron star. (This is consistent with an
analysis of the photon spectral break in the GeV regime, -- see Ray et al. 
1999).

As shown in Table 1, PSR1706-44 has similar $P$ and $\dot P$ to those of
the Vela pulsar. It is therefore of interest that also the Vela pulsar is 
faint in the optical with $V=23.65$ and $A_{\rm V} \lsim 0.4$ (Nasuti et al.
1997). If PSR1706-44 would have the same optical luminosity as Vela, it could 
have a $V$ magnitude of $V = 26.9$ (if its $A_{\rm V}$ is $0.9$ magnitudes, 
and its distance is 1.8 kpc). Our non-detection of PSR1706-44 in $V$ is 
consistent with this. However, theory indicates that the optical emission is
sensitive to many parameters, so there is certainly room for PSR1706-44 to be
intrinsically much brighter than the Vela pulsar. This is also consistent with
our results, in particular if PSR1706-44 lies close to Star 1 in projection.

\begin{acknowledgements}
We thank the Science Verification team for the observation of this pulsar and
making available the summed image for analysis, and Claes-Ingvar Bj\"ornsson
for discussions. P.L. and J.S. are supported by the Swedish National Space 
Board, and P.L. is also supported by the Swedish Natural Science Research 
Council. The work of A.R. is part of the 9th Five Year Plan project 9P-208[a] 
at Tata Institute.
\end{acknowledgements}

{}
  

\begin{thebibliography}{}

\bibitem [Becker et al. 1995] {BBT} Becker, W., Brazier, K.T.S., Tr\"umper, J.
1995, A \& A,  298, 528

\bibitem [Chakrabarty \& Kaspi 1998] {CK98} Chakrabarty, D., Kaspi, V.M. 1998,
ApJ, 498, L37 (CK98)

\bibitem [Cheng, Ho \& Ruderman 1986a] {CHR86a} Cheng, K.S., Ho, C.,
Ruderman, M.A. 1986, ApJ, 300, 500

\bibitem [Cheng, Ho \& Ruderman 1986b] {CHR86b} Cheng, K.S., Ho, C.,
Ruderman, M.A. 1986, ApJ, 300, 522

\bibitem [Finley et al. 1998] {Fin98} Finley, J.P., Srinivasan, R., Saito, Y.
et al. 1998, ApJ, 493, 884

\bibitem [Frail \& Goss 1998] {FG98} Frail, D.A., Goss, W.M. 1998,
in preparation

\bibitem [Goldoni et al. 1995] {GMCB95} Goldoni, P., Musso, C., Caraveo, P.A.,
Bignami, G.F. 1995, A \& A, 298, 535.

\bibitem [Gull et al. 1998] {Gu98} Gull, T.R., Lindler, D.J., Crenshaw, D.M.
et al. 1998, ApJ, 495, L51

\bibitem [Koribalski et al. 1995] {Ko95} Koribalski, B., Johnston, S., 
Weisberg, J.M. \& Wilson, W. 1995, ApJ, 441, 756

\bibitem [Nasuti et al. 1997] {Na97} Nasuti, F.P., Mignani, R., Caraveo, P.A.,
Bignami, G.F. 1997, A \& A, 323, 839

\bibitem [Pacini \& Salvati 1987] {PS87} Pacini, F. \& Salvati, M. 1987,
ApJ, 321, 447

\bibitem [Ray, Harding \& Strickman 1999] {RHS99} Ray, A., Harding, A.K.,
Strickman, M.S. 1999, ApJ, March 10 issue. 

\bibitem [Spitzer 1978] {S78} Spitzer, L. 1978, Physical Processes in the 
Interstellar Medium (New York: Wiley) 

\bibitem [Taylor et al. 1993] {T93} Taylor, J.H., Manchester, R.N. \& Lyne,
A.G. 1993, ApJS, 88, 529

\bibitem [Thompson et al. 1996] {T96} Thompson, D. 1996, in Pulsars: 
Problems \& Progress (eds. S. Johnston et al.) Astron. Soc. Pacific Ser. 105,
307

\bibitem [Thompson et al. 1996] {T96} Thompson, D.J., Bailes, M., Bertsch, D.L.
et al. 1996, ApJ, 385, 465

\bibitem [Thompson et al. 1999] {T98} Thompson, D.J., Bailes, M., Bertsch, D.L.
1999, ApJ, in press (astro-ph/9811219)

\bibitem [Usov 1994] {U94} Usov, V.V. 1994, ApJ 427, 394

\bibitem [Wang et al. 1998] {Wa98} Wang, N., Manchester, R.N., Bailes, M. et 
al. 1998, MNRAS, submitted


\end{thebibliography}
\end{document}